\documentstyle[12pt]{article}

\def\be{\begin{equation}}
\def\ee{\end{equation}}
\def\bea{\begin{eqnarray}}
\def\eea{\end{eqnarray}}
\def\ov{\overline}

\def\nnu{\nonumber}
\def\ll{\label}

\begin{document}
\newcommand{\Psl}{\not\!\! P}
\newcommand{\dsl}{\not\! \partial}
\newcommand{\half}{{\textstyle\frac{1}{2}}}
\newcommand{\for}{{\textstyle\frac{1}{4}}}
\newcommand{\eqn}[1]{(\ref{#1})}
\newcommand{\npb}[3]{ {\bf Nucl. Phys. B}{#1} ({#2}) {#3}}
\newcommand{\pr}[3]{ {\bf Phys. Rep. }{#1} ({#2}) {#3}}
\newcommand{\plb}[3]{ {\bf Phys. Lett. B}{#1} ({#2}) {#3}}
\newcommand{\prd}[3]{ {\bf Phys. Rev. D}{#1} ({#2}) {#3}}
\newcommand{\hepth}[1]{ [{\bf hep-th}/{#1}]}
\newcommand{\grqc}[1]{ [{\bf gr-qc}/{#1}]}

\def\la{\mathrel{\mathpalette\fun <}}
\def\a{\alpha}
\def\b{\beta}
\def\g{\gamma}\def\G{\Gamma}
\def\d{\delta}\def\D{\Delta}
\def\e{\epsilon}
\def\et{\eta}
\def\z{\zeta}
\def\t{\theta}\def\T{\Theta}
\def\l{\lambda}\def\L{\Lambda}
\def\m{\mu}
\def\f{\phi}\def\F{\Phi}
\def\n{\nu}
\def\p{\psi}\def\P{\Psi}
\def\r{\rho}
\def\s{\sigma}\def\S{\Sigma}
\def\ta{\tau}
\def\x{\chi}
\def\o{\omega}\def\O{\Omega}
\def\lagr{{\cal L}}
\def\cd{{\cal D}}
\def\k{\kappa}
\def\be{\begin{equation}}
\def\ee{\end{equation}}
\def\tz{\tilde z}
\def\tF{\tilde F}
\def\ri {\rightarrow}
\def\cf{{\cal F}}
\def\pa {\partial}
\def\bpa{\overline{\pa}}
\def\ov{\over}
\begin{flushright}
DFPD/97/TH/48\\ hep-th/9710189\\
October, 1997
\end{flushright}
\bigskip\bigskip
\begin{center}
{\large\bf MACROSCOPIC STRING-LIKE SOLUTIONS IN MASSIVE SUPERGRAVITY
}
\footnote{ This work is supported by INFN fellowship}

\vskip .9 cm
{\sc Harvendra Singh}
\footnote{e-mail: hsingh@pd.infn.it} 
 \vskip 0.05cm
I.N.F.N. Sezione di Padova, Departimento di Fisica ``Galileo Galilei'', \\
Via F. Marzolo-8, 35131 Padova, Italy
\end{center}
\bigskip
\centerline{\bf ABSTRACT}
\bigskip
\begin{quote}
In this report we obtain explicit string-like solutions of
 equations of motion of massive heterotic supergravity 
recently obtained by Bergshoeff, Roo and Eyras. 
We also find consistent string source which can be embedded 
in these backgrounds when   space-time 
dimension is greater than or equal to six.

\end{quote}

\newpage

Recently, there has been regenerated interest in the study of massive 
supergravity theories
\cite{berg,cow,bre,ms} 
and the generalised Scherk and Schwarz dimensional reduction scheme 
\cite{ss,berg}.
In generalised Scherk-Schwarz (GSS) toroidal reduction some of the fields 
are given linear dependence along the coordinates of the torus.
Thus the 
resultant compactified theory in the lower dimensions possesses mass like 
parameters. However, when these parameters are set to zero the massive theory 
reduces to the `standard' (massless) supergravity. Another point to be 
imphasized here is that the massive supergravities obtained 
through GSS  
reduction, in general, possess smaller duality symmetry groups
than their massless
counterparts \cite{ms,bre}. 

Fundamental string solution  
\cite{dh} was obtained in the spacetimes which are
asymptotically Minkowskian. Later on other 
fundamental solutions (with source terms)   
as well as solitonic  solutions
(without source terms) were obtained for any p-brane in D spacetime
dimensions \cite{du}. For these solutions  masses and their respective 
charges saturate the
Bogomol'nyi-Prasad-Sommerfeld bound and therefore the supersymmetry 
in the theory
dictates that these classical solutions are the quantum mechanically 
exact solutions of 
the theory. It is now natural to ask
what does happen if the spacetime around  a p-brane is not
asymptotically flat or if there is nontrivial dilatonic 
potential in the theory. 
Such examples are  provided when we consider massive supergravities, 
$e.g.$, massive IIA supergravity in $D=10$ \cite{roma} and its subsequent 
dimensional reductions\cite{berg,ms}, various GSS reductions of massless 
type II \cite{berg,cow} and recently of 
heterotic strings in ten dimensions \cite{bre}. All these massive theories
have some kind of dilatonic potentials either in the NS-NS sector as 
is the case 
with massive heterotic of Bergshoeff, Roo and Eyras or in the R-R sector 
as was the case with massive type IIA of 
Romans \cite{roma,cow} or in both the sectors as was anticipated in 
\cite{ms} in order to obtain maximally symmetric black hole solutions
analogous to \cite{sv}.

In the present work, first, we shall obtain
explicit string like solutions of
source free string equations of motion  in
arbitrary spacetime dimensions with nontrivial dilatonic potential 
of the form suggessted in the massive heterotic string \cite{bre}. In these
solutions spacetime has 
explicit $O(1,1)\times O(D-2)$ symmetry and does not have
asymptotic flatness. Secondly, we shall show that a string like source
could be consistently embedded in these spacetimes  when 
$D\ge 6$. 

We consider the following  effective action in D-dimensional
target space,
\be
S= \int d^Dx \sqrt{-g} e^{-2\F} \bigg[ R_g + 4 \ \pa_\mu\F \pa^\mu\F
-{1\over 2\cdot3!} H_{\mu\nu\l} H^{\mu\nu\l} - 2 m^2 \bigg]
\label{11}
\ee
where $m^2$ is the mass term (or cosmological constant) as in \cite{bre} 
or an analog of 
central charge deficit term, $\F$ is the dilaton
field and $g_{\mu\nu}$ is the  $\sigma-$model metric.
We have taken $2 \kappa^2 ~=~ 1 ~=~2\pi\alpha'$.
The antisymmtric field strength is
expressed as,
\be
H_{\mu\nu\l} = \pa_\mu B_{\nu\l} + {\rm cyclic \ permutations}.
\label{12}
\ee
Above action for $D=4$ can be obtained from the appropriate truncation 
of the massive theory described in \cite{bre}.
Since it would be convenient to work  in the Einstein (canonical) frame let
us write down the action in the canonical metric, $ g_{\mu\nu}=
e^{{a\over 2} \f} G_{\mu\nu}$
\be
S= \int d^Dx \sqrt{-G} \bigg[ R_G -{1\over2} \pa_\mu \f\pa^\mu\f
-{1\over 2.3!} e^{ -a \f} H_{\mu\nu\l} H^{\mu\nu\l} - 2 m^2
e^{{4\over a (D-2)}\f}\bigg]
\label{13}
\ee
where $ a=\sqrt{ 8\ov D-2}$ and rescaled dilaton is $\F ={\f\ov
a}$. One can derive equations of motion from the action \eqn{13}. We obtain 
the solutions to these equations in two cases; when $B_{\m\n}\ne 0$ and when
$B_{\m\n}=0$. The latter solutions we describe as the vacuum solution (pure 
dilaton gravity). 

\noindent{\bf Case-I $B_{\m\n}\ne 0$}

We find   that the equations of motion derived from \eqn{13} are satisfied for
the following choice of the background fields
\bea
&&ds^2= U^{-(D-4)\ov(D-2)}( -dt^2 + dx^2) + U^{2\ov(D-2)}( dy_1^2
+ \cdots + dy_{D-2}^2),\nnu \\ 
&&\f= -{a\ov 2} \ln U,\nnu \\ 
&&B= {1\ov 2} B_{\m\n} dx^\m \wedge dx^\n = -{1\ov U} dx^0 dx^1,
\label{14}
\eea
provided the potential has the following form
\bea
U= 1+ {m^2\ov 2}(t^2 - x^2) + {Q_2 \ov |y-y_0|^{D-4}}, \ \ \ \
{\rm for}\ D>4 \nnu \\ 
= 1+ {m^2 \ov 2}(t^2 -x^2) - Q_2 \ln |y-y_0|, \ \ \ \  D=4
\label{15}
\eea
where $ Q_2$ is the charge associated with the 2-from gauge
field $B_{\m\n}$  and is given by
\be
Q_2= \int_{S^{D-3}} e^{-a \f}\ {}^\ast H.
\ll{19}\ee
The symbol $\ast$ stands for Hodge dual operation and
 $y_0$ is some point in the transverse
y-plane. 

Obviously, the background solution obtained in \eqn{14} is similar
to the  spacetime  of a
fundamental string solution in \cite{dh}. Only difference is of an
explicit $m$-dependent term in \eqn{15}. In the limit $m \to 0$ 
background in \eqn{14} reduces to
asymptotically flat spacetime around a fundamental string.
There is a curvature singularity at $y=y_0$ 
which can be smoothened by introducing some string-like source
at $y=y_0$.  

\noindent{\bf Case-II $B_{\m\n}=0$}

In this case we have a pure dilatonic gravity with the dilatonic potential.
The equations of motion are still 
satisfied by the same ansatz 
as in \eqn{14} 
for the remaining fields while
 the potential in \eqn{15} becomes
\bea
U= 1+ {m^2\ov 2}(t^2 - x^2), \ \ \ \ {\rm for}~ D > 2.
\label{15a}
\eea
Thus we also have solution independent of the antisymmetric field strength.
When this background is substituted into the action it becomes
\be
S=\int d^D x \bigg[ {4 m^2 \ov D-2}\bigg],
\ee
which shows there is finite lagrangian density ${4 m^2 \ov D-2}$ per unit 
$D$-dimensional 
Minkowski volume. 

We now introduce string-like source in the form of $\s$-model world-sheet
action
\be
S_\s= -{1\ov 2} \int d\ta d\s \bigg[ \sqrt{-\g} \g^{ij}
 \pa_iX^M\pa_jX^N G_{M N} e^{{a \ov 2}\f} +
\epsilon^{ij}\pa_iX^M\pa_jX^N B_{M N}\bigg]
\label{16}
\ee
where $\g_{ij}$ is the induced metric on the string world-sheet. The 
$\f$-dependence is chosen in accordance with \cite{du} so that, when
$m \to 0$, under the rescaling
\bea
&&G_{MN}\to \l^{4\ov D-2} G_{MN},~~~~~~ B_{MN}\to\l^2B_{MN}\nnu\\
&&e^\f\to\l^{4(D-4)\ov(D-2)a} e^\f,~~~~~\g_{ij}\to\l^2\g_{ij}
\label{16a}
\eea
both actions $S$ and $S_\s$ scale in the same way
\be
S\to\l^2 S,~~~~~~S_\s\to\l^2 S_\s.
\label{16b}
\ee
When $m\ne 0$ such rescaling of the action does not hold, see \cite{gm}.
Now, the above string like
source can be embedded in the spacetime  \eqn{14} if we
make the following choice of static gauge for the world-sheet coordinates,
\bea
&&X^0=\ta, \ \ X^1=\s, \ \ X^r= {\rm constant}, \ \ \ \
(r=1,...,D-2), \nnu\\
&&\g_{ij}=\pa_iX^M\pa_jX^N G_{M N} e^{{a\ov2}\f}.
\label{17}
\eea
That is to say, eqs.(\ref{14}) and (\ref{17}) together satisfy
all the field equations derived from actions $S$ and $ S_\s$
if considered together at least for $D\ge 6$. The essential 
requirement for the embedding of the source  \eqn{16} 
is that the string coupling $e^\f$ vanishes at $y=y_0$, $i.e.$
\be
U_{y\to y_0}\to \infty.
\ll{18}\ee

For $D< 6$ above condition \eqn{18} is violated near the end points
 of the string,
$i.e.$, when  $x\to\pm\infty$. Therefore we have difficulty to find 
 appropriate source when $D <6$. However, if the end points are excluded,
which is possible for temporal evolution $(t > x)$, 
then the source action in \eqn{16}
is a good choice even for $D< 6$. Thus in \eqn{14} we 
have got a background configuration in which a charged macroscopic 
string is embedded in a cosmological  spacetime. 
These solutions have explicit
$O(1,1)\times O(D-2)$ symmetry. The symmetry gets automatically enhanced to 
$P_2\times O(D-2)$ when $m\to 0$. Here $P_2$ stands for P\"oncare symmetry
in two dimensions. This constitutes our main result.
  
\par Next we shall try to find multi-string solution in asymptotically
non-flat spacetime $\eqn{14}$. First we calculate the net transverse force a
test string is subjected to  when another string is brought close to it. 
We still have \cite{dh}
\be
{d^2\ov d\ta^2} X^m = 2 \G^m_{0 0} + H^m_{0 0} = 0,
\label{21}
\ee
where $\G^\m_{ \n \s}$ are the Christoffel connections in the
canonical metric $G_{\m\n}$. The result in eq. \eqn{21}  suggests that net
transverse force between two such strings vanishes. Thus it can
be argued that a multi-string solutions could also be obtained in
this configuration.
For multi-string case the potential $U$ in \eqn{15} modifies to
\bea
U= 1+ {m^2\ov 2}(t^2 - x^2) + \sum_n{Q_2 \ov |y-y_n|^{D-4}}, \ \ \ \
{\rm for}\ D>4 \nnu\\
= 1+ {m^2 \ov 2}(t^2 -x^2) - Q_2 \sum_n \ln |y-y_n|, \ \ \ \  D=4
\label{22}
\eea
where $n$ is the number of strings fixed at the positions $y_n$
in the y-plane. It is not easy to calculate ADM energy for
these solutions. We do not know whether their masses and respective 
charges will 
saturate the BPS bound as they do in  the case when $m=0$. However, since 
 we can control the value of the mass parameter $m$ it appears to us
that for $m\sim 0$ 
the solutions obtained above will only be slightly off extremal. 


\par To analyse briefly the evolution and spacetime 
properties of the metric
in (\ref{14}) we consider specific case of $D=6$ and of single
string source positioned at $y=0$. The metric is
\be
ds^2= U^{-1\ov2}\bigg[( -dt^2 + dx^2) + U \ ( dy_1^2
+ \cdots + dy_4^2) \bigg], 
\label{31}\ee
with $U= 1+ {m^2\ov 2}(t^2 - x^2) + {Q_2 \ov y^2}$. The spacetime
is non-static and the evolution is dragged by a Lorentz
invariant quantity involving the directions tengential to the
string world-sheet. 
The evolution could be {\it light-like} if the quantity $ (t^2 -x^2)$ 
is greater than zero and it could be also {\it space-like} if $(t^2-x^2)$
is less than zero. The spacetime can be viewed as a solenoid whose axis is
identified with the x-direction. 
 At any given time
the metric in \eqn{31} is not well defind in whole space since
U becomes negative whenever $ 1+ {m^2 t^2 / 2} + {Q_2/y^2} < m^2 
x^2 /2$.  However, a light signal travelling in x-direction will not see any
discontinuity and only will detect the geometry around a 
fundamental string straightened at $y=0$.
For a possible $m^2 < 0$ (anti-de-Sitter) case, situation is more interesting.
There is a critical time $t_c =\sqrt{2\ov |m^2|}$ such that for $t<t_c$ the
metric is well defind in the whole space except at $y=0$ which
is the position of the string source. As $|m^2| \to 0$ we can see
that $t_c\to \infty$. This implies that for infinitesimally small values of
the cosmological constant, $or$ for a slow rate of evolution,
it will take infinitely longer time for any irregularities to
set in in the space. But $m^2$ cannot become negative if it is the mass 
term comming from GSS reduction \cite{bre}. 
Certainly there must be some other source
for negative cosmological constant in the NS-NS sector 
if that has to happen. 

I have been  benifited by some useful discussions with Prof. M. Tonin and 
specially with 
Prof. J. Maharana. The beginning part of this work was carried out at Institue
of Physics, Bhubaneswar.


\begin{thebibliography}{99}
\bibitem{berg} E. Bergshoeff, M. de Roo, M. Green, G. 
Papadopoulos and P. Townsend,
 \npb{470}{1996}{113},\hepth{9601150};\\
 E. Bergshoeff and M. B.
Green, {\it The type IIA super-eight brane}, preprint VG-12/95,\\
M.B. Green, C.M. Hull and P.K. Townsend, \hepth{9604119}.
\bibitem{cow} P. Cowdall, H. Lu, C. N. Pope, K.S. Stelle and P. K. Townsend,
{\bf Nucl. Phys. B}486 (1997) 49,\\
 I. V. Lavrinneko, H. Lu and C. N. Pope, {\it From topology to
generalised dimensional reduction}, CTP-TAMU-59/96, \hepth{9611134},\\
E. Bergshoeff, P.M. Cowdall and P.K. Townsend, {\it Massive type IIA 
supergravity from the topologically massive D-2-brane}, preprint UG-6/97, 
\hepth{9707139}.
\bibitem{bre} E. Bergshoeff, M. de Roo and E. Eyras, {\it Gauged supergravity
from dimensional reduction}, preprint UG-7/97, \hepth{9707130}.
\bibitem{ms} J. Maharana and H. Singh, {\it On the compactification of 
type II string theory}, \plb{408}{1997}{164}, 
for typo corrected version see \hepth{9505058}.
\bibitem{ss} J. Scherk and J. H. Schwarz, {\bf Nucl. Phys. B}153 (1979) 61; 
{\bf Phys.
Lett. B}82 (1979) 60.
\bibitem{dh} A. Dabholkar, G. Gibbons, J. Harvey and F.
Ruiz-Ruiz, \npb{340}{1990}{53}.
\bibitem{du} M. Duff, R. Khuri and J. Lu, \pr{259}{1995}{213}.
\bibitem{roma} L.J. Romans, {\bf Phys. Lett. B}169 (1986) 374.
\bibitem{sv} A. Strominger and C. Vafa, \plb{379}{1996}{99}, \hepth{961029}.
\bibitem{gm}G. Gibbons and K. Maeda, \npb{298}{1988}{741}.
\end{thebibliography}
\end{document}